\documentclass[twocolumn,epsbox]{jpsj2}
\newcounter{u-large}
\newcounter{u'-large}
\def\runtitle{
Anomalous  Flux Quantization in  Hubbard Ring 
}
\def\runauthor{Kazuhiro {\sc Sano} and Yoshiaki {\sc \=Ono} }

\newcommand{\simj}{\stackrel{>}{_\sim}}

\title{%
Anomalous  Flux Quantization in  the Spin-Imbalanced Attractive  Hubbard Ring}

\author{%
Kazuhiro {\sc Sano,}\thanks{E-mail address: sano@phen.mie-u.ac.jp}\raisebox{0.5ex}{1}  and Yoshiaki {\sc \=Ono}\raisebox{0.5ex}{2}  
}

\inst{%
\raisebox{0.5ex}{1}Department of Physics Engineering,  Mie University, Tsu, Mie 514-8507, Japan       \\ 
\raisebox{0.5ex}{2}Department of Physics, Niigata University, Ikarashi, Nishi-ku, Niigata 950-21817, Japan  
}

\recdate{\today}

\abst{
We investigate the one-dimensional Hubbard ring with attractive interaction in the presence of imbalanced spin populations by using the exact diagonalization method. The singlet pairing correlation function is found to show spatial oscillations with power-law decay as expected in the Fulde-Ferrell-Larkin-Ovchinnikov state of a Tomonaga-Luttinger liquid. 
In the strong coupling regime, the system shows an anomalous flux quantization of period $h/4e$,  half of the superconducting flux quantum of $h/2e$, as recently predicted by  mean-field analysis, together with various flux quanta smaller than $h/4e$. Notably, the observed flux quanta are determined by the difference between the system size $N_L$ and  electron number $N_e$ as $h/(N_L-N_e)e$. 
}

\kword{%
 FFLO state, flux quantization, spin imbalance, attractive Hubbard model,  exact diagonalization
}

\begin{document}
\sloppy
\maketitle

\section{Introduction}

Ultracold atomic gases in optical traps have stimulated much interest in  fundamental problems of  superfluid states\cite{Partridge,Lewensteina,Giorgini,Liao}.  In particular, two component Fermi particle systems  with different populations are attracting much attention due to the possibility of an exotic superfluid phase, which is the so-called Fulde-Ferrell-Larkin-Ovchinnikov (FFLO) state\cite{Fulde,Larkin}. 
The FFLO state is characterized by the formation of Cooper pairs with finite center-of-mass momentum caused by the imbalance of the Fermi surfaces of two-component fermions and exhibits inhomogeneous superconducting phases with a spatially oscillating order parameter.

Recent theoretical works have revealed that the FFLO state accompanies a variety of spatial structures of the pairing order parameter, which is dominated by the difference between the Fermi wave vectors of the two component fermions.\cite{Matsuda,Yanase1,Yanase2,Meisner,Lutchyn,Rizzi,Quan,Moreo}
On the basis of the bosonization approach and   conformal field theory, L\"uscher {\it et al.} have discussed the correlation exponents of the FFLO state in the one-dimensional (1D) attractive Hubbard model and have obtained a phase diagram of the electronic states on the parameter plane of  electron density vs. the spin imbalance.\cite{Luscher} This diagram indicates that the  FFLO region  expands with increasing  strength of the attractive interaction. 

More recently, Yoshida and Yanase\cite{Yoshida} have pointed out that the FFLO state in a mesoscopic ring with 200 sites exhibits an anomalous  flux quantization of  period $h/4e$, which is   half of the superconducting flux quantum $h/2e$. 
They  solved the 1D attractive Hubbard model on the basis of the mean-field approximation in the weak-coupling BCS region with the use of the Bogoliubov-de Gennes equation\cite{Yoshida}. However, the strong quantum fluctuation effect, which is considered to be crucial for the 1D mesoscopic ring, together with the strong coupling effect in the  BCS to Bose-Einstein condensation (BEC) crossover region, was not  discussed. 
To achieve a certain understanding  including such effects beyond the mean-field  approximation, we believe that  nonperturbative and reliable approaches  are required. 

In the present paper,  we investigate the  1D Hubbard ring with the attractive interaction by using the exact diagonalization (ED) method for finite-size systems. To confirm the  anomalous  flux quantization of period $h/4e$, we numerically calculate the periodicity of the ground-state energy $E(\Phi)$ with respect to the magnetic flux $\Phi$ without any approximation. 
Although our calculation is restricted to small  systems, we  expect that the essential features of the  FFLO state can be well described even in  finite-size systems, as previously discussed by several authors,\cite{Meisner,Lutchyn,Rizzi,Quan,Moreo,Luscher} in the case with $\Phi=0$. 

\section{Model and Formulation}
We consider the 1D  Hubbard  ring given by the following Hamiltonian:
\begin{eqnarray} 
 H=-t\sum_{i,\sigma}(e^{i2\pi \Phi/N_{L}} c_{i,\sigma}^{\dagger} c_{i+1,\sigma}+h.c.) 
  -|U|\sum_{i}n_{i,\uparrow}n_{i,\downarrow},  
\label{hund-Hamil}  
 \end{eqnarray} 
where $c^{\dagger}_{i,\sigma}$ stands for the creation operator for an
 electron with spin $\sigma \ (= \uparrow, \downarrow)$  at site $i$  and 
   $n_{i,\sigma}=c_{i,\sigma}^{\dagger}c_{i,\sigma}$.
Here, $t$ represents the hopping integral between nearest-neighbor sites, and we set $t=1$ in this study.  $\Phi$ corresponds to the magnetic flux through the ring
measured in units of $h/e$, and $N_{L}$ is the system size. The interaction parameter $|U|$ stands for the strength of the attractive interaction on the site.

We numerically diagonalize the  model Hamiltonian [eq. (\ref{hund-Hamil})] of up to 20 sites  using the standard Lanczos algorithm.
To carry out a systematic calculation, we use the periodic boundary  condition when $N_\uparrow$ and $N_\downarrow$  are odd numbers and the antiperiodic boundary condition when they are even, where $N_\uparrow$ and $N_\downarrow$ are the total numbers of up- and down-spin electrons, respectively\cite{Ogata}. 
 The filling $n$ of electrons is given  by  $n=N_{e}/N_{L}$, where $N_e(=N_\uparrow + N_\downarrow)$ is the total number of electrons, and the spin imbalance is defined by $p=\frac{N_\uparrow -N_\downarrow}{N_e}$.
We also calculate the correlation function $C(r)$ of singlet superconducting pairing  at the same site as 
\begin{eqnarray}
  C(r)=\frac1{N_L}\sum_{i} \langle c^\dagger_{i,\uparrow}
 c^\dagger_{i,\downarrow} c_{i+r,\downarrow} c_{i+r,\uparrow}\rangle.     
\end{eqnarray}
\section{Pairing Correlation}
\begin{figure}[t]
\begin{center}
\includegraphics[width=6.0cm]{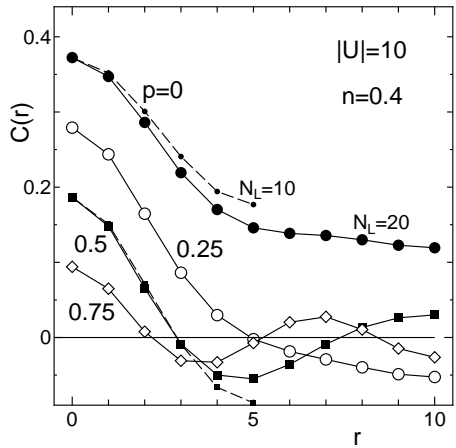}
\end{center}
\caption{Singlet pairing correlation  functions  $C(r)$  as functions of $r$ for several values of $p$  at $n$=0.4 for $|U|=10$. The solid lines represent the results for  $N_{L}=20$ and the dashed lines represent those for $N_{L}=10$.}
\label{sc-cor}
\end{figure}

To confirm  the FFLO states  and  estimate the finite-size effect in   small  systems, we examine the singlet pairing correlation functions $C(r)$ for two different-size systems,  $N_{L}=10$ and $N_{L}=20$.  
In Fig. \ref{sc-cor}, we show  $C(r)$ as functions of $r$ for  $p=$0 and 0.5 with $N_{L}=10$ and  for $p=$0, 0.25, 0.5, and 0.75 with $N_{L}=20$. 
Comparing $C(r)$ for $p=$0 and 0.5 between  $N_{L}=10$ and 20, we find that the discrepancy between the two sizes is not so large. Thus, the finite-size effect is expected to be small even in  small  systems with $N_{L}\simj 10$.\cite{finite-effect}

%
%%%%%%%%%%%%%  SC correlation   C(r)     %%%%%%%%%%%%%%%%%%%%%%%%%%%%%%
%
%%%%%%%%%%%%%%%%%%%%%%%%%%%%%%%%%%%%%%%%%%%%%%%%%%%%%%%%%%%%%%%%%%%%%%%

%%%%%%%%%%%%%  SC correlation   C(r)     %%%%%%%%%%%%%%%%%%%%%%%%%%%%%%
%
\begin{figure}[t]
\begin{center}
\includegraphics[width=6.0cm]{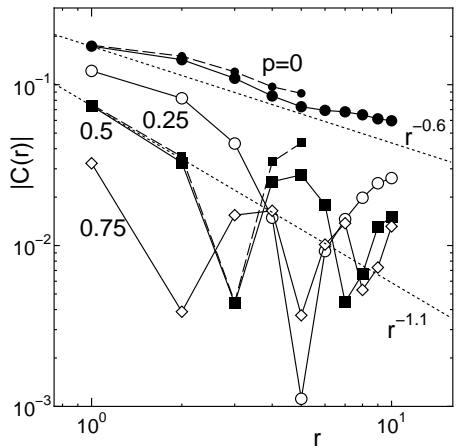}
\end{center}
\caption{Logarithmic plot of $|C(r)|$  as functions of $r$  for several values of $p$ at $n$=0.4 for $|U|=10$. The solid lines represent the results for  $N_{L}=20$ and the dashed lines represent those for $N_{L}=10$. The dotted lines are guides for the eyes. }
\label{logplot}
\end{figure}
%%%%%%%%%%%%%%%%%%%%%%%%%%%%%%%%%%%%%%%%%%%%%%%%%%%%%%%%%%%%%%%%%%%%%%%

We find that $C(r)$ for $p=0$  decays monotonically as a function of $r$, while  $C(r)$ for $p\ne 0$  shows oscillating behavior. The period of the oscillation seems to decrease  with increasing  $p$.
This result is related to the spatial oscillation of the superconducting order parameter $\Delta(r)$ which has already been  pointed out in  previous works.\cite{Matsuda,Yanase1,Yanase2,Meisner,Lutchyn,Rizzi,Quan,Moreo,Luscher,Yoshida} 
The periodicity of $\Delta(r)$ is determined by the difference between the Fermi wave numbers of $k_{F\uparrow}$ and $k_{F\downarrow}$  and varies as $\Delta(r) \propto \exp(iQr)$ for the Fulde-Ferrell state and as  $\Delta(r) \propto \cos(Qr)$ for the Larkin-Ovchinnikov state, where $Q=|k_{F\uparrow}-k_{F\downarrow}|=n\pi p$.\cite{Luscher}
In Fig. \ref{sc-cor}, we see that  the spatial period of $C(r)$ with $n=0.4$  is 20, 10, and 20/3 for $p=0.25$, 0.5, and 0.75, where the corresponding values of $Q=n\pi p$ are $\pi/10$, $\pi/5$, and $3\pi/10$, respectively. Thus, we can conclude that the oscillating behavior of $C(r)$ is well accounted for by the FFLO state  similar to the case of $\Delta (r)$\cite{Matsuda,Yanase1,Yanase2,Meisner,Lutchyn,Rizzi,Quan,Moreo,Luscher,Yoshida} mentioned above. 

Note that, in the purely 1D system, the correlation function $C(r)$ should show power-law decay  for large $r$, as predicted by  Tomonaga-Luttinger liquid theory,\cite{Solyom,Voit,Yin} instead of  superconducting long-range order with finite $\Delta(r)$ obtained from the mean-field approximation.  
As shown in Fig. \ref{logplot}, the logarithmic plot of $|C(r)|$   indicates that  the pair correlation  function  decays as $\sim r^{-0.6}$ for $p=0$, and  as $\sim r^{-1.1}$ for $p>0$, in addition to the oscillating behavior mentioned before. The present ED results with $N_{L}=10$ and $N_{L}=20$ seem to be  consistent with the results obtained from the density-matrix renormalization group\cite{Luscher} for a larger  system with $N_L=64$.\cite{c-decay}
This suggests that the characteristic feature of the FFLO state of a Tomonaga-Luttinger liquid is well described even in  small  systems with $N_{L}\simj 10$. 

\section{Flux Quantization}
Next, we discuss the flux quantization of the  FFLO state in the attractive Hubbard ring.  
Recently, Yoshida and Yanase have shown that an  anomalous flux quantization  with a period of $h/4e$, which is  half of the superconducting flux quantum $h/2e$, occurs for the  FFLO state  in  mesoscopic rings.\cite{Yoshida}
They claimed that the FFLO  state has a multicomponent order parameter and shows rich phases distinguished by the order parameter  in the presence of the mass flow, which corresponds  to  the magnetic flux through the ring in our model.
When  the  mass flow increases, the FFLO state undergoes a transition to another type of FFLO state with a different order parameter.  
Therefore, the ground state of our system is expected to alternate between  several types of FFLO states with increasing of the flux, resulting in  anomalous flux quantization such as the half period $h/4e$. 
To verify this prediction without any approximation, we calculate the periodicity of the ground-state energy $E(\Phi)$ with respect to the magnetic flux $\Phi$ by using the ED method.\cite{Kusakabe} Although the system is limited to small sizes, this should give  direct evidence of the anomalous flux quantization of the FFLO state, which is well described even in  small  systems with $N_{L}\simj 10$ as shown in \S 3.

%
%%%%%%%%%%%%%  flux  n8e4 U10,200,1000        %%%%%%%%%%%%%%%%%%%%%%%%%%
%
\begin{figure}[t]
\begin{center}
\includegraphics[width=5.2cm]{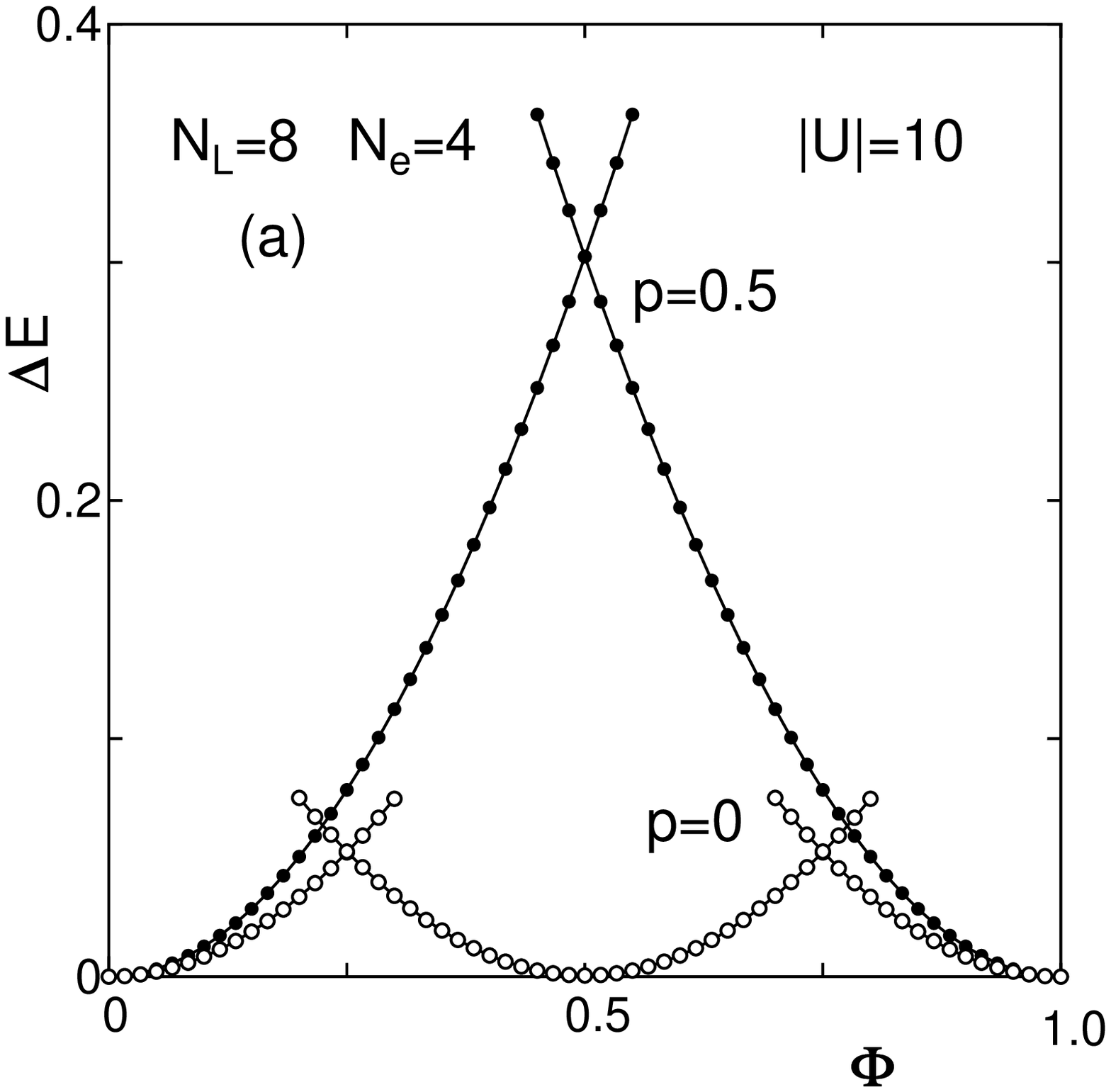}
\includegraphics[width=5.2cm]{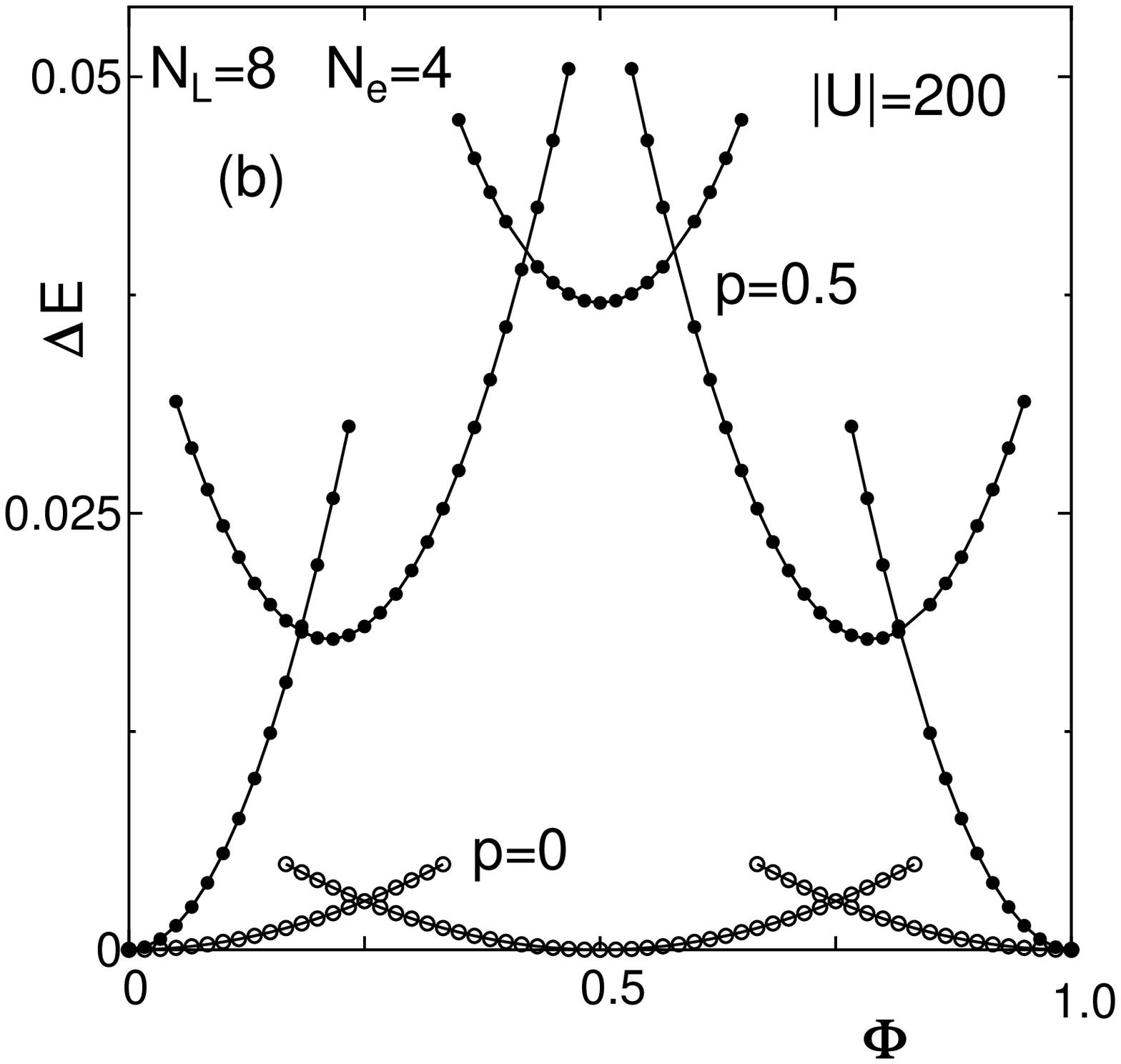}
\includegraphics[width=5.2cm]{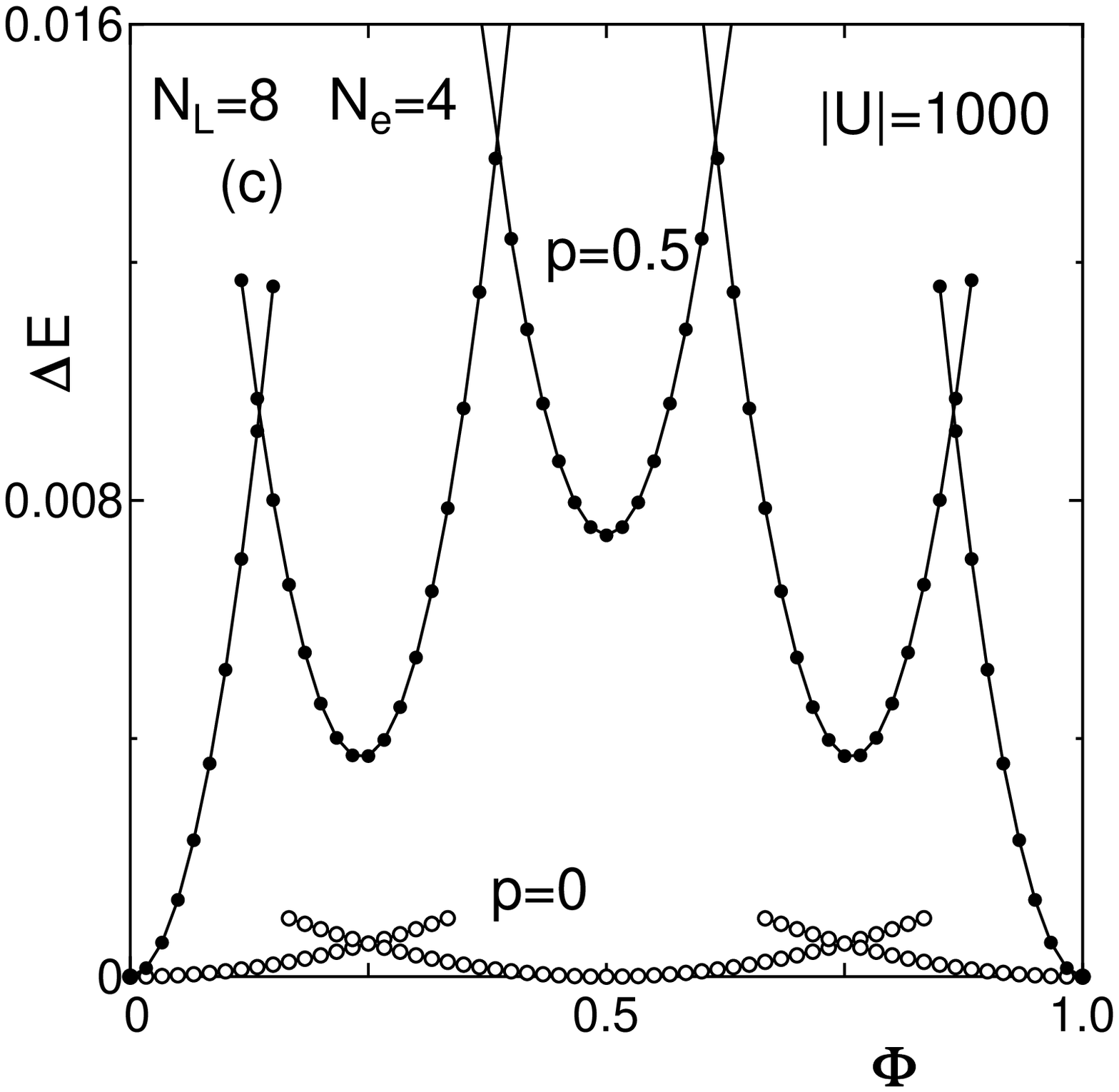}
\end{center}
\caption{Difference in the ground-state energy $\Delta E=E(\Phi)-E(0)$ as a function of the  magnetic flux $\Phi$ with $p=0$ and 0.5 at $n$=0.5 (4 electrons/8 sites) for $|U|=10$ (a),  $|U|=200$ (b), and  $|U|=1000$ (c).}
\label{flux-8site}
\end{figure}
%%%%%%%%%%%%%%%%%%%%%%%%%%%%%%%%%%%%%%%%%%%%%%%%%%%%%%%%%%%%%%%%%%%%%%%

%
%%%%%%%%%%%%%  flux  n8e4,16e8 U2000        %%%%%%%%%%%%%%%%%%%%%%%%%%
%
\begin{figure}[t]
\begin{center}
\includegraphics[width=6.2cm]{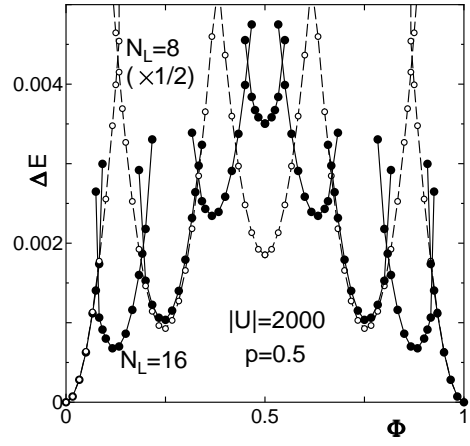}
\end{center}
\caption{Difference  in the ground-state energy $\Delta E=E(\Phi)-E(0)$ as a function of the magnetic flux $\Phi$ with $p=0.5$  at $n$=0.5 (4 electrons/8 sites and  8 electrons/16 sites)  for  $|U|=2000$.}
\label{flux-16site}
\end{figure} 
%%%%%%%%%%%%%%%%%%%%%%%%%%%%%%%%%%%%%%%%%%%%%%%%%%%%%%%%%%%%%%%%%%%%%%%
%
%%%%%%%%%%%%%%%%%%%%%%%%%%%%%%%%%%%%%%%%%%%%%%%%%%%%%%%%%%%%%%%%%%%%%%%
%
\begin{figure}[t]
\begin{center}
\includegraphics[width=6.7cm]{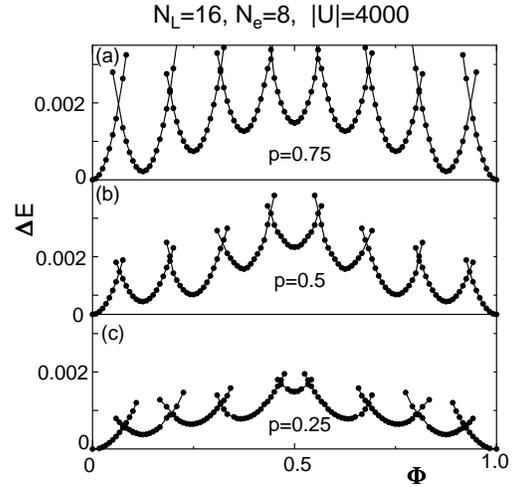}
\end{center}
\caption{Difference in the ground-state energy $\Delta E=E(\Phi)-E(0)$ as a function of the magnetic flux $\Phi$ at $n=0.5$ (8 electrons/16 sites) for $|U|=4000$ with $p=0.75$ (a), $p=0.5$ (b),  and $p=0.25$ (c).}
\label{ne-fix}
\end{figure}
%%%%%%%%%%%%%%%%%%%%%%%%%%%%%%%%%%%%%%%%%%%%%%%%%%%%%%%%%%%%%%%%%%%%%%%

%%%%%%%%%%%%%%%%%%%%%%%%%%%%%%%%%%%%%%%%%%%%%%%%%%%%%%%%%%%%%%%%%%%%%%%
%
\begin{figure}[t]
\begin{center}
\includegraphics[width=6.7cm]{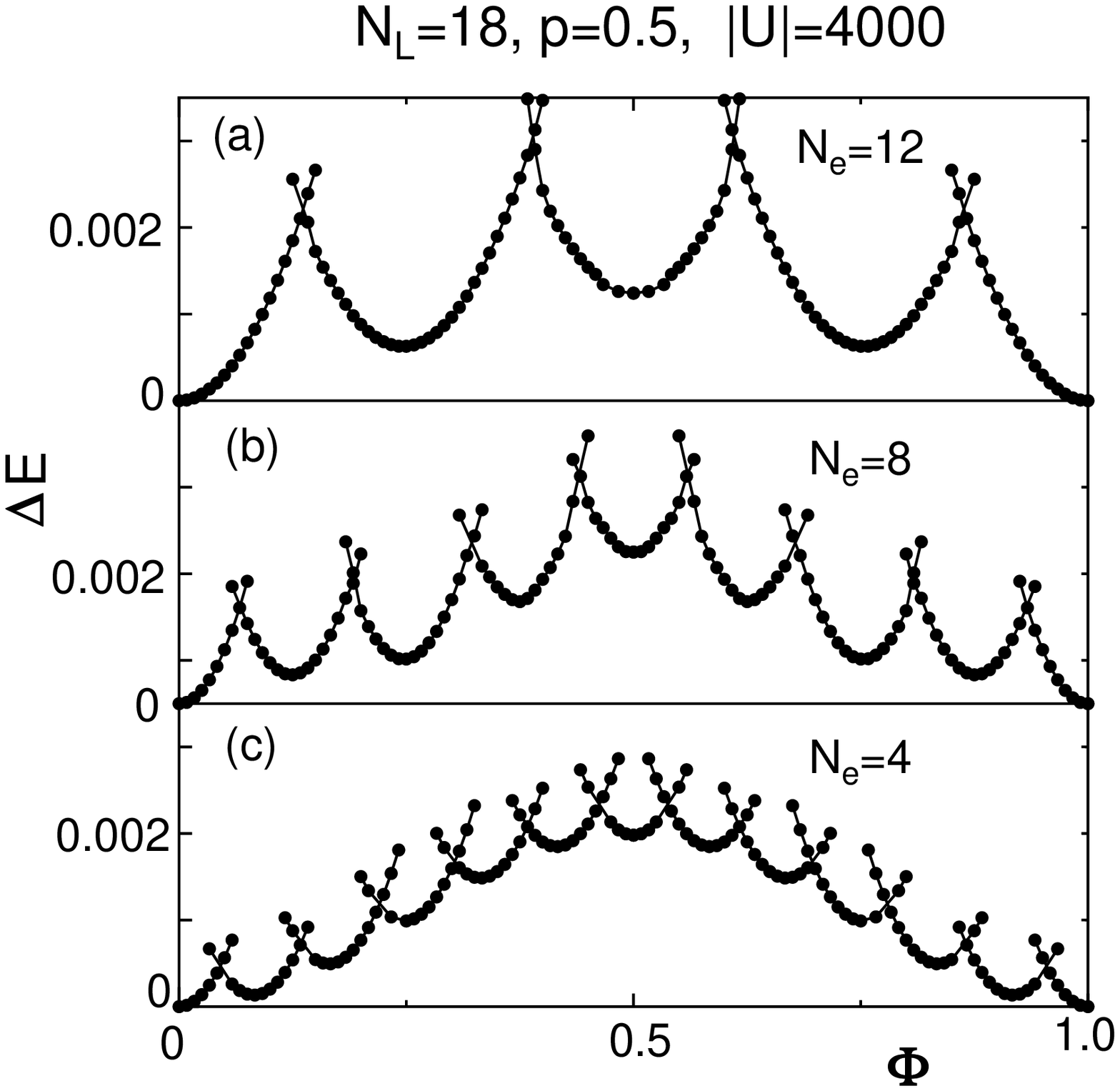}
\end{center}
\caption{Difference in the ground-state energy $\Delta E=E(\Phi)-E(0)$ as a function of the magnetic flux $\Phi$ with  $p=0.5$ at $N_L=18$ and $|U|=4000$ for  $N_e=12$ (a), $N_e=8$ (b),  and $N_4=4$ (c).}
\label{p-fix}
\end{figure}
%%%%%%%%%%%%%%%%%%%%%%%%%%%%%%%%%%%%%%%%%%%%%%%%%%%%%%%%%%%%%%%%%%%%%%%

 In Figs. \ref{flux-8site}(a)-\ref{flux-8site}(c), we show the difference in the ground-state energy $\Delta E(\Phi)=E(\Phi)-E(0)$  as a function of  $\Phi$  for quarter filling $n$=0.5 (4 electrons/8 sites)  at  $|U|=10$, 200,  and  1000, respectively.
In the absence of spin imbalance, i.e.,  $p=0$, we see that the energy levels cross  at  $\Phi \sim 0.25$ and 0.75  for all values of the attractive interaction $|U|$, where the usual superconducting flux quantization of period $h/2e$ ($=0.5$ in the present unit of $h/e=1$) is observed. 
On the other hand, in the presence of  spin-imbalance with $p=0.5$, the energy levels cross  at  $\Phi \sim 0.5$ for a relatively small value of $|U|=10$, as shown in Fig. \ref{flux-8site}(a), while they cross at  $\Phi \sim 0.2$, 0.4, 0.6, and 0.8 for relatively large values of $|U|=200$  and  1000 as shown in Figs. \ref{flux-8site}(b) and \ref{flux-8site}(c), respectively, where the period of the ground-state energy with respect to $\Phi$ for $p=0.5$ is clearly  half of that for $p=0$; thus, the anomalous  flux quantization of  period $h/4e$ takes place. 
The obtained result of the anomalous  flux quantization seems to agree with that obtained from the mean-field approximation.\cite{Yoshida} 
However, the anomalous  flux quantization is observed exclusively for relatively large $|U|$ in the present study, while it is observed for small $|U|$ in the mean-field study.\cite{Yoshida}
This discrepancy will be discussed in \S 5.

In Fig. \ref{flux-16site}, we show a comparison of $\Delta E(\Phi)$  between two different-size systems,  $N_L=8$ and 16, with $p=0.5$ at $n=0.5$ for $|U|=2000$. 
Here we plot $\frac{\Delta E}{2}$ for $N_L=8$ for comparison with $\Delta E$ for $N_L=16$ and find that both coincide well with each other for small $\Phi$. The result suggests that  the size dependence of $\Delta E$ is approximately given by $1/N_L$ for small $\Phi$. 
If we assume  $E(\Phi)$ to be an analytic function with respect to $\Phi$,  $E(\Phi)$ can be expanded in a power series of $\Phi$ as 

\begin{eqnarray}
&E(\Phi)&\!\!\!\!\!=E(0) \nonumber \\
&+&\!\!\!\!\!\!\!\!\!\!
\left. \frac{1}{2!}\frac{\partial^2 E(\Phi)}{\partial \Phi^2}\right |_{\phi=0} \Phi^2+
\left. \frac{1}{4!}\frac{\partial^4 E(\Phi)}{\partial \Phi^4}\right |_{\phi=0} \Phi^4+... , \ \ \ \ \ \label{EPhi}
\end{eqnarray}
where  the odd powers of $\Phi$  vanish because of the inversion symmetry, $E(\Phi)=E(-\Phi)$. 
Then, eq. (\ref{EPhi}) yields for small $\Phi$
\begin{eqnarray}
  \Delta E(\Phi) \simeq 2\pi D\Phi^2/N_L,
\end{eqnarray}
where $D$ is the Drude weight;\cite{Khon,Fye}  $D= \frac{N_L}{4\pi} \left. \frac{\partial^2 E(\Phi)}{\partial \Phi^2}\right|_{\Phi=0}$ and is expected to be independent of $N_L$ for large $N_L$. 
This indicates that  the  size dependence of $\Delta E(\Phi)$  is dominated  by $1/N_L$.
Therefore, it is considered to be difficult to observe the flux quantization for large $N_L$ since the value of $\Delta E(\Phi)$ is small for large $N_L$ and finally becomes zero in the limit $N_L \to \infty$. 
In fact,   Yoshida and Yanase\cite{Yoshida} claimed that the anomalous  flux quantization of $h/4e$ is observed only in the case with the mesoscopic ring and that it disappears in the limit $N_L \to \infty$. 

In Fig. \ref{flux-16site}, we also observe an anomalous flux quantization with a period of $h/8e$, a quarter of the superconducting flux quantum $h/2e$, for the $N_L=16$ system, in contrast to the $N_L=8$ system where a flux quantization with a period of $h/4e$ is observed. 
It is very curious that the two systems exhibit  different flux quanta even though both  have the same $n$  and $p$, and only the system size $N_L$ is different.
To clarify this puzzling behavior, we perform detailed calculations for various systems. 
Figure \ref{ne-fix} shows the $\Phi$ dependence of $\Delta E$ for  $p=0.25$, 0.5, and 0.75 with $N_L=16$ and $N_e=8$. 
It clearly indicates that the period of these systems is the same regardless of $p$ and is given by $h/8e$.  In Fig. \ref{p-fix}, we plot the $\Phi$ dependence of $\Delta E$ for several values of $N_e$ with $N_L=18$ and $p=0.5$, and we find that the flux quanta are  $h/4e$ for  $N_e=12$,   $h/8e$ for  $N_e=8$, and $h/12e$ for  $N_e=4$. These results suggest that the flux quanta of the anomalous flux quantization exclusively depend on  $N_L$ and  $N_e$ and are independent of $p$ except when $p=0$ (usual superconducting state with the flux quantum $h/2e$) and $p=1$ (fully spin-polarized noninteracting system where the periodicity of $\Delta E(\Phi)$ is $h/e$). 

\begin{figure}[t]
\begin{center}
\includegraphics[width=7.2cm]{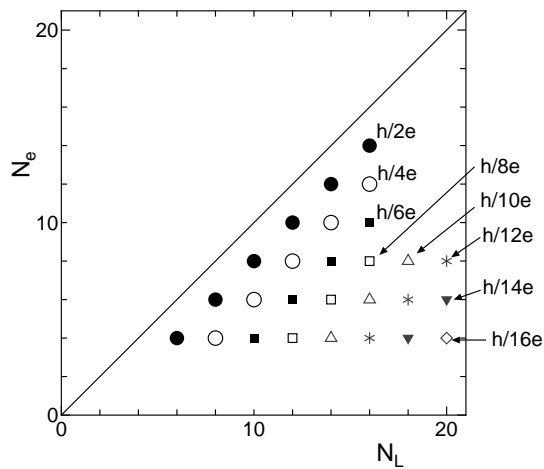}
\end{center}
\caption{Phase diagram  of the observed flux quanta  on the $N_L$-$N_e$ plane  with $0<p<1$ for sufficiently large $|U|$.}
\label{souzu}
\end{figure}

We perform systematic calculations of the $\Phi$ dependence of $\Delta E$ for available systems to determine the flux quanta observed for sufficiently large $|U|$ and obtain the phase diagram on the $N_L$-$N_e$ plane shown in Fig. \ref{souzu}, where we confirm that the flux quanta are independent of $p$ for $0<p<1$ as mentioned before. 
Notably, the observed flux quanta are determined by the difference between the system size $N_L$ and  electron number $N_e$ as $h/(N_L-N_e)e$ as shown in Fig. \ref{souzu}.\cite{fig7-odd}  
When we apply the result to large systems, we obtain various types of anomalous flux quantization including tiny flux quanta in the case of large  $N_L-N_e$. 
This is in striking contrast to the case with the mean-field approximation, where  half flux quantum $h/4e$ is exclusively observed for the mesoscopic ring with $N_L=200$.\cite{Yoshida}
This discrepancy will be discussed in \S 5.

%%%%%%%%%%%%%%%%%%%%%%%%%%%%%%%%%%%%%%%%%%%%%%%%%%%%%%%%%%%%%%%%%%%%%%%
%
%%%%%%%%%%%%%%%%%%%%%%%%%%%%%%%%%%%%%%%%%%%%%%%%%%%%%%%%%%%%%%%%%%%%%%%

To qualitatively understand the origin of the anomalous flux quantization with the notable flux quanta $h/(N_L-N_e)e$, let us consider the attractive Hubbard ring in the strong coupling limit $|U| \to \infty$ for simplicity. 
In this limit, the systems with $0<p<1$, i.e., $N_\uparrow>N_\downarrow >0$, consist of  $N_\downarrow$ electron pairs on the same sites (doublons) and  $N_\uparrow-N_\downarrow$ unpaired up-spin electrons. 
When a doublon moves to the nearest-neighbor site through the hopping of a down-spin electron of the doublon, an unpaired up-spin electron must be at the nearest-neighbor site so as not to lose the binding energy $-|U|$. 
Therefore, the doublons move in the ring accompanied by the unpaired up-spin electrons as explicitly shown below.

Figure \ref{e-hopping2} shows a schematic diagram of the possible hopping process of a doublon  to the nearest-neighbor site for the simplest system with $N_e=3$ ($N_\uparrow=2$, $N_\downarrow=1$) consisting of a doublon and an unpaired up-spin electron in the attractive Hubbard ring with $N_L$ sites in the limit $|U| \to \infty$. 
First, we assume the initial state where the doublon is at the first site and  the up-spin electron is at the second site. 
Next, we move the up-spin electron from  the second site to the $N_L$th site through $N_L-2$ counterclockwise hops. 
Finally, we move the down-spin electron of the doublon from the first site to the $N_L$th site through one  clockwise hops to obtain the final state where the doublon is at the $N_L$th site and  the up-spin electron is at the first site. 
Then the configuration of the final state is rotated in the clockwise direction by one site from the initial state.

The above-mentioned process contains $N_L-2$  counterclockwise electron hops,  and one  clockwise electron hop, corresponding to the $N_L-3$  counterclockwise electron hops. Thus, $E(\Phi)$ is considered to exhibit  oscillation with a period of $h/(N_L-3)e$ as a single electron hops under the magnetic flux $\Phi$ through the ring is responsible for the oscillation of $E(\Phi)$ with the period of $h/e$. 
This is in striking contrast to the case of the usual superconducting state with $p=0$, where a doublon (or a Cooper pair) hopping process contains two  electron hops, resulting in the oscillation of $E(\Phi)$ with the period of $h/2e$, that is, the superconducting flux quantum. 
For the general case with $N_e$ electrons where $N_\uparrow>N_\downarrow > 0$, a similar rotation process contains $N_L-N_\uparrow$  counterclockwise electron hops  and $N_\downarrow$  clockwise electron hops, corresponding to  $N_L-N_e$   counterclockwise electron hops. Thus, $E(\Phi)$ is considered to show  oscillation with a period of $h/(N_L-N_e)e$, which is precisely the anomalous flux quantum observed in the present study.

%%%%%%%%%%%  electron hopping process        %%%%%%%%%%%%%%%%%%%%%%%%%%
%
\begin{figure}[t]
\begin{center}
\includegraphics[width=5.0cm]{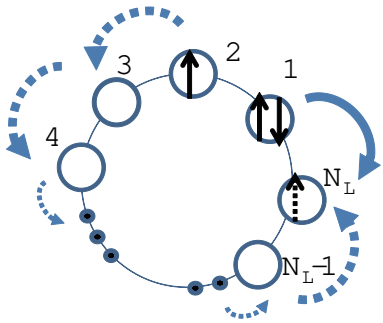}
\end{center}
\caption{(Color online) Possible  hopping process of a doublon to the nearest-neighbor site for the simplest case with $N_e=3$ ($N_\uparrow=2$, $N_\downarrow=1$)  in the attractive Hubbard ring with $N_L$ sites in the limit $|U| \to \infty$.
}
\label{e-hopping2}
\end{figure}
%%%%%%%%%%%%%%%%%%%%%%%%%%%%%%%%%%%%%%%%%%%%%%%%%%%%%%%%%%%%%%%%%%%%%%%

%
% ;;;;;;;;;;;;;   Summary and Discussion  ;;;;;;;;;;;;;;;;;;;;;;;;;
% 
 \section{Summary and Discussion}
We have investigated the 1D attractive Hubbard ring in the presence of  spin imbalance  by using the ED method, which enables us to obtain reliable results including the strong coupling regime. 
We have found the following. 

(1) The singlet pairing correlation functions show spatial oscillations with power-law decays as expected in the FFLO state of a Tomonaga-Luttinger liquid.
 
(2) In the strong coupling regime, the system shows an anomalous flux quantization with the flux quantum $h/4e$, as recently obtained by Yoshida and Yanase\cite{Yoshida} in a  mean-field study, together with various flux quanta smaller than $h/4e$.
 
(3) The observed flux quanta are determined by the difference between the system size $N_L$ and  electron number $N_e$ as $h/(N_L-N_e)e$, independent of the spin imbalance except when $p=0$ and $p=1$.

Although our ED calculations are restricted to small systems of up to $N_L=20$,  the analysis in the strong coupling limit $|U| \to \infty$ has confirmed the anomalous flux quantization with the notable flux quanta $h/(N_L-N_e)e$. 
Similar behavior has already been discussed in the repulsive ($U>0$) Hubbard ring by using the Bethe ansatz in the strong correlation regime with $U\gg n|t|$, where $E(\Phi)$ shows  oscillation with a period of $h/N_e$.\cite{Kusmartsev1,Kusmartsev2} 
However, by using the well-known transformation of the attractive Hubbard model into the repulsive one,\cite{Luscher,Emery,Woynarovich} both systems with the anomalous flux quantization are found to be  different from each other.\cite{U-transformation} 
An interesting future problem will be to study the anomalous flux quantization in the attractive Hubbard ring for larger  systems by using the Bethe ansatz and to elucidate the finite-size effect as it might be responsible for the stability of the anomalous flux quantization, which is observed exclusively for the strong coupling regime ($|U|\simj 100$) in the present ED calculations but for the weak coupling regime ($|U|\sim 1.5$) in the mean-field approximation.\cite{Yoshida}

Finally, we compare  the present results and the previous mean-field results concerning the flux quantum. 
Various flux quanta given by $h/(N_L-N_e)e$ are observed for the 1D attractive Hubbard ring in the present study, while only  half flux quantum $h/4e$ was observed in the mean-field study\cite{Yoshida} where three-dimensionality is implicitly included. Therefore, the present results of the various flux quanta except $h/4e$ might disappear when we include the effects of three-dimensionality. 
In addition, although the mean-field approximation is appropriate for the weak coupling regime, the correlation effects, which are neglected in the approximation, become important for the strong coupling regime, where the various flux quanta are observed in the present study. 
Thus, the flux quanta except $h/4e$ might be observed with  the correlation effects even in the presence of the three dimensionality in the strong coupling regime. 
To state our findings more  conclusively, it is necessary to  investigate  the 1D attractive Hubbard ring  including the effects of three-dimensionality together with the use of the Bethe ansatz available for larger systems.

\section*{Acknowledgments}
The authors thank Y. Yanase and T. Yoshida for directing our attention to the problem addressed in this study and for useful comments and discussions.  This work was partially supported by a Grant-in-Aid for Scientific Research from the Ministry of Education, Culture,  Sports, Science  and Technology.


\begin{thebibliography}{99}
\bibitem{finite-effect} 
The absolute value of $C(r)$ for $N_L=10$ is slightly  larger than that for  $N_L=20$.
This suggests that  the finite-size effect enhances the pairing correlation.
\bibitem{c-decay}
The exponent of  $C(r)$ seems to be slightly larger than the result of ref. \ref{Luscher}  because of the finite-size effect.
\bibitem{fig7-odd}
Although the case of $N_L$ and/or $N_e$ being an odd number is not shown in this figure, we have confirmed that the same result is obtained.
\bibitem{U-transformation}
If we had applied the transformation to our system irrationally, we would have obtained the incorrect periodicity of  $h/(N_L-pN_e)e$.
\end{thebibliography}
\end{document}